\documentclass[aps,prl,twocolumn,superscriptaddress]{revtex4}
\usepackage{amsmath,amssymb,graphicx}
\usepackage{amsbsy}
\usepackage{latexsym}
\usepackage{color}
\usepackage{graphicx}
\usepackage{psfrag}
\usepackage[normalem]{ulem}
\usepackage{bm}
\usepackage{url}
\usepackage{notes2bib}
\usepackage{soul}

\newcommand{\be}{\begin{equation}}
\newcommand{\ee}{\end{equation}}
\newcommand{\bea}{\begin{eqnarray}}
\newcommand{\eea}{\end{eqnarray}}

\newcommand{\comment}[1]{}

\begin{document}

\title{Maximal Diversity and Zipf's Law}

\author{Onofrio Mazzarisi}
\email{mazzaris@mis.mpg.de}
\affiliation{Max Planck Institute for Mathematics in the Sciences, Inselstraße 22, 04103 Leipzig, Germany}
\affiliation{Dipartimento di Fisica ``E.~R. Caianiello'',Universit\`a  di Salerno, via Giovanni Paolo II 132, 84084 Fisciano (SA), Italy}

\author{Amanda de Azevedo-Lopes}
\affiliation{Instituto de F\'\i sica, Universidade Federal do Rio Grande do Sul, CP 15051, 91501-970 Porto Alegre RS, Brazil}

\author{Jeferson J. Arenzon}
\affiliation{Instituto de F\'\i sica, Universidade Federal do Rio Grande do Sul, CP 15051, 91501-970 Porto Alegre RS, Brazil}
\affiliation{Instituto Nacional de Ci\^encia e Tecnologia - Sistemas Complexos, Rio de Janeiro RJ, Brazil}

\author{Federico Corberi}
\affiliation{Dipartimento di Fisica ``E.~R. Caianiello'',Universit\`a  di Salerno, via Giovanni Paolo II 132, 84084 Fisciano (SA), Italy}
\affiliation{INFN, Gruppo Collegato di Salerno,
and CNISM, Unit\`a di Salerno, Universit\`a  di Salerno, via Giovanni Paolo II 132, 84084 Fisciano (SA), Italy}


\begin{abstract}
Zipf's law describes the empirical size distribution of the components of many systems
in natural and social sciences and humanities.
We show, by solving a statistical model, that
Zipf's law co-occurs with the maximization of 
the diversity of the component sizes.
The law ruling the increase of such diversity
with the total dimension of the system is derived and its relation with Heaps' law is discussed.
As an example, we show that our analytical results compare very well with linguistics and population datasets.
\end{abstract}

\maketitle

Diversity is a central concept in ecology, economics, information theory, and other natural and social sciences.
It can be quantified by diversity indices~\cite{Jost06,Tuomisto10}, 
such as (species) richness, the Gini-Simpson index or Boltzmann-Shannon entropy,
which characterize the system under study from different angles.
Loosely understanding the term,  
high diversity may represent an advantage in terms of resilience and performance. 
This is the case, for instance, in ecology, where well differentiated ecosystems are often 
(see, e.g., Ref.~\cite{Ives58} for the debate on this topic) considered to be more stable~\cite{Elton1958,Tilman06,AreseLucini20}, 
and in economy as well: strong countries have a well diversified production~\cite{Tacchella12}. 

In most cases diversity is hindered by limiting factors.
For an ecosystem the amount of energy and chemical components available
does not allow an unbounded increase of the population.
Similarly, the number of different items produced by an economy is limited by its strength. 
The diversity drift is therefore a complex optimization process.

Elaborating on that, in this Letter we 
take the aforementioned restrictions into account and, 
among the possible measures of diversity~\cite{Jost06} 
we consider the richness index $D$, 
which turns out to be  particularly suited for a quantitative description of such optimization tendency
in many complex systems.
Richness is
a quantity that counts the number 
of different types which are present in a collection of items. 
For instance, the set of integers 
$\{3,7,1,9,0,1\}$ is richer than $\{3,2,3,7,7,2\}$, 
because there are 5 different figures in the former and only 3 in the latter.
Every diversity measure can be rephrased in terms of 
R\'enyi~\cite{renyi1961measures} (or, equivalently, Tsallis~\cite{Tsallis88}) entropies (see Ref.~\cite{Jost06}
and Supplemental Material (SM)~\cite{Sup}). 
Notice, however, that the index $D$ alone is insensitive to the abundance of each type but only to their
presence/absence.

We consider situations where types can be identified by quantitative labels $s$, as in the example above. 
$D$ is the richness of the collection of entities $\{s_1,\dots,s_N\}$, with arbitrary $N$, but subjected
to the additive constraint $S=\sum_{n=1}^Ns_n$. Here $s_n$ represents the portion of the total resource $S$
assigned to the $n$-th entity of the ensemble, i.e. its size.
Entities can be cities~\cite{Gabaix99} of a country with total population $S$,
distinct words~\cite{Piantadosi14} occurring with absolute frequencies $\{s_n\}$ in a book of size $S$
or genes~\cite{Furusawa03} expressed with abundances $\{s_n\}$ where $S$ is the total number of proteins synthesized in a cell.

These systems are instances where the Zipf's law~\cite{Zipf49,Newman05} is observed to hold.
Other well known examples include~\cite{Clauset09} GDP of nations~\cite{Cristelli12},
firm sizes~\cite{Axtell01}, species in taxa~\cite{Willis22} and fragmentation processes~\cite{OdDiBo93}. 
If ranked according to their size $s$, components obey Zipf's law when
\begin{equation}
s(r)\propto r^{-a},
\label{zipf}
\end{equation}
where $r$ is the rank, with $a \simeq 1$.
A representation in terms of the distribution of sizes~\cite{Newman05,Corral} $p(s)\propto s^{-\tau}$, with 
$\tau = 1+a^{-1}$, is better suited to our purposes.
To explain Zipfian behavior many generative mechanisms have been 
proposed~\cite{Simon55,Levy96,Marsili98,Cancho03,TrLoSeSt14,Corominas15,Mazzolini18}
and it has also been framed in a broader statistical perspective~\cite{Mora11,Marsili13,Schwab14}.
For instance, it has been shown to be associated to
maximally informative samples
in modeling complex systems~\cite{Marsili13,Cubero_2019}.

In this Letter we show that the maximization of the diversity index $D$
and the occurrence of Zipf's law in the distribution of the component sizes $\{s_n\}$
are naturally related.
This is achieved by deriving, in a statistical model, a {\it diversity law} that
can be used to estimate the index $D$ of distributions of empirical data.
We put our results to the test showing 
remarkable agreement with data for quantitative linguistics, taken from the Gutenberg English texts database~\cite{ProjectGutemberg}, and for urbanistics from the GeoNames database~\cite{GeoNames}.
Finally, within our approach we also recover in a simple way the expression of Heaps' law~\cite{Heaps78,LuZhZh10}
and discuss its relation with the diversity law.
The fact that specifically $D$, among the possible diversity measures,
is extremized,
indicates the prominent role played by this quantity in the 
many and diverse natural phenomena described by the Zipf's law
and represents a different and perhaps profitable rationalization for its occurrence.

{\it The model.---}Consider sets of independent and identically distributed integer random variables $\{s_n\}$, sampled
from a generic probability distribution $p(s)$. 
We call $s_n$ the size of the $n$-th component (or entity).
$p(s)$ will be denoted as the
{\it bare} distribution, since the effective ({\it dressed}) distribution of the $s_n$ is shaped by the presence of a global constraint $\sum _{n=1}^N s_n=S$,
where $S$  is the total dimension of the system.
$N$ is the fluctuating number
of entities that, according to the particular extraction of the $\{s_n\}$, is needed to fulfill the
constraint.
The probability of a particular configuration 
\(\mathcal{C}\equiv [\{s_1,\ldots,s_N\};N]\) is given by
\begin{equation}
\label{eq:formal_conditioned}
p_S(\{s_1,\ldots,s_N\};N)=\frac{1}{Z_S}\prod_{n=1}^{N}p(s_n)\delta_{\sum_{n=1}^Ns_n,S} \ ,
\end{equation}
where the constraint is enforced by the Kronecker delta.
The quantities $Z_S=\sum_{N=1}^{\infty}Z_S(N)$ and
\begin{equation}
\label{eq:partial_part}
Z_S(N)\equiv\sum_{s_1=1}^{S}\sum_{s_2=1}^{S}...\sum_{s_N=1}^{S}\prod_{n=1}^{N}
p(s_n)\delta_{\sum_{n=1}^Ns_n,S}\,\,\,
\end{equation}
play the role of partition functions in an ensemble where $N$ is fluctuating or fixed, respectively.
One obtains the probability of having a number $N$ of entities as $p_S(N)=Z_S(N)/Z_S$.
The dressed probability of observing a size $s$ can be written using Eq.~(\ref{eq:partial_part}) as
\begin{equation}
\label{eq:conditioned_explicit}
p_S(s)=\frac{p(s)}{\sum_{N=1}^{\infty}N \ Z_S(N)}\sum_{N=1}^{\infty}N \ Z_{S-s}(N-1) \ ,
\end{equation}
where the factor \(N\) appears because we do not distinguish among components.

If $t_{s}$ is the number of times the value $s\in [1,S]$ is found in a given configuration ${\cal C}$,
the diversity index $D$ (hereafter also referred to as simply {\it diversity}) is defined as
\begin{equation}
D = \sum_{s=1}^S \left( 1-\delta_{t_{s},0} \right),
\end{equation}
namely the number of different values assumed by the entities. 
The probability $p_S(D)$ of observing a certain value of $D$ is formally given in
the SM~\cite{Sup}.

\begin{figure}
\includegraphics[width=0.75\linewidth]{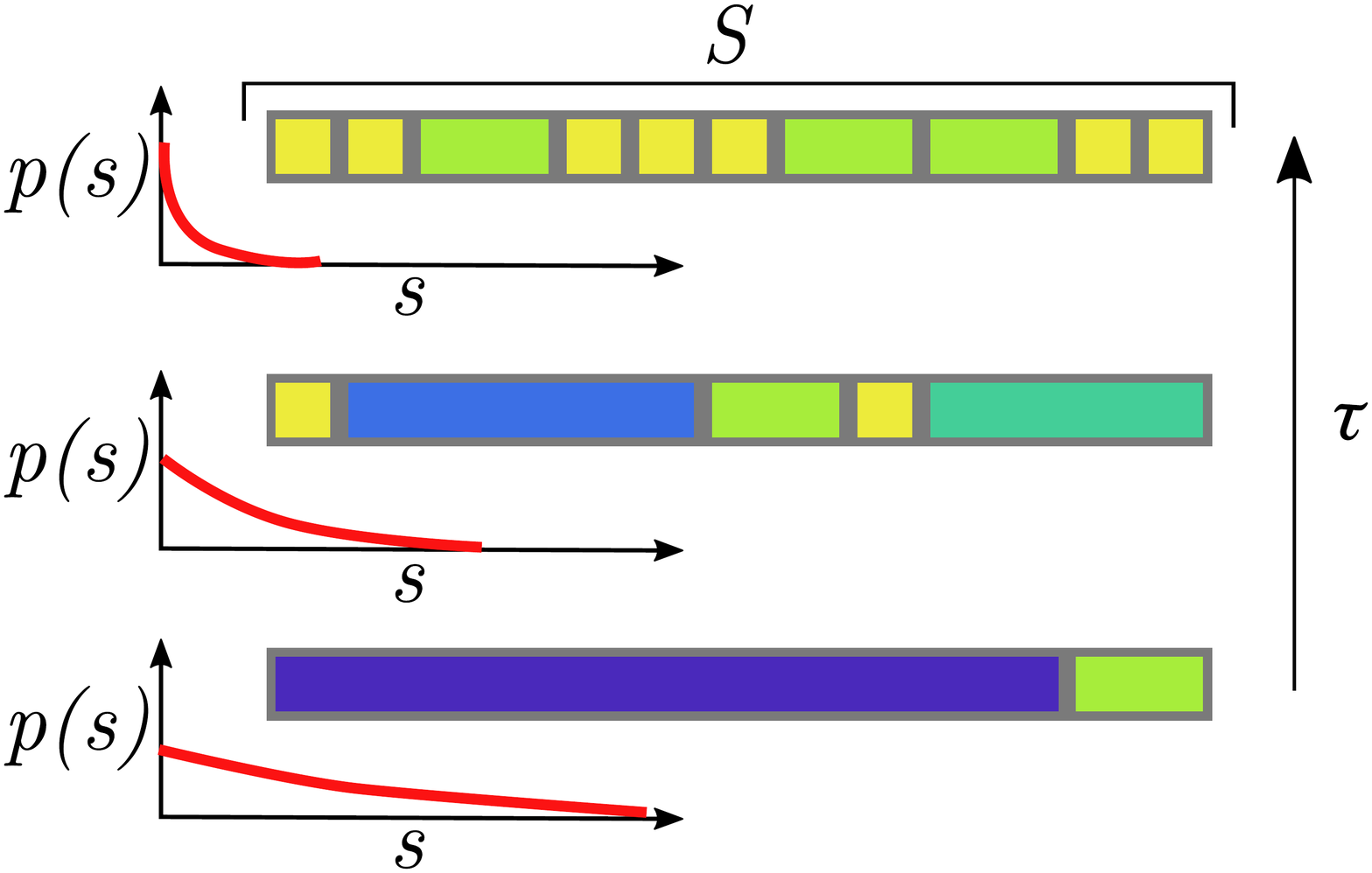}
\caption{Pictorial representation of the problem. Power laws, $p(s)\sim s^{-\tau}$, are sketched with an increasing exponent $\tau$
(from bottom to top) alongside with relative typical realizations $\{s_n\}$. Entities of the same size
are depicted as blocks of the same color and in all the cases they add up to $S$, the total length of the bar.
For large values of $\tau$, most of the entities have small and similar sizes, resulting in a poor 
diversity $D$. In the other limit, small $\tau$,  large
sizes do get more probable but the total number of entities required to fill $S$ is smaller. Consequently,
diversity is again small.
The diversity is expected to be maximal for an intermediate value of $\tau$.}
\label{pic:pictorial}
\end{figure}

We are interested in highly diverse configurations, therefore we consider
power law bare probability distributions, which grant access to a wide range of sizes,
\begin{equation}
p(s)=\frac{s^{-\tau}}{\Lambda(\tau, S)} \quad; \, \mbox{for}\quad  1\leq s\le S,
\label{barealg}
\end{equation}
and $p(s)=0$ otherwise. The normalization $\Lambda (\tau,S)=\zeta(\tau)-\zeta(\tau,S+1)$ is a generalized harmonic number and
can be written in terms of the Riemann and Hurwitz zeta functions,
$\zeta(x)$ and $\zeta(x,y)$ respectively.

Our goal is to compute the average diversity $\langle D\rangle_S$ and the value of $\tau$
which maximizes it (see Fig.~\ref{pic:pictorial}).
Given the complicated expression of $p_{S}(D)$, 
we directly determine $\langle D\rangle _S$ as follows.
We split the range of sizes into $s\le s^*$ and $s>s^*$~\cite{AzRoOlAr20}, 
where $s^*$ is defined by $\langle N\rangle_S \,\,p_S(s^*)=1$;
these two sectors contribute to $\langle D\rangle _S$ as
\begin{equation}
\langle D\rangle_S\simeq s^*+\langle N\rangle _S\sum _{s=s^*}^S p_S(s).
\label{JefH}
\end{equation}
Indeed, given an average number 
of entities $\langle N\rangle_S$, there is at least one of them for each size 
$s\le s^*$, contributing to the first term on the r.h.s. of Eq.~(\ref{JefH}).
The second term is the average number of entities with $s>s^*$.
Since these are represented at most once this also corresponds to their contribution to $\langle D\rangle_S$.

With Eq.~(\ref{JefH}), the evaluation of $\langle D\rangle _S$ only depends on 
the knowledge of $\langle N\rangle _S$ and $p_S(s)$. 
These quantities can be computed numerically with an exact recursive method, 
as discussed in the SM~\cite{Sup}.
For an analytical treatment of the problem 
it is possible to approximate the dressed probability distribution
with the bare one, i.e. $p_S(s)\simeq p(s)$ (see the SM~\cite{Sup}).
This simplification leads to an asymptotic expression for $\langle D\rangle _S$ which is accurate for large $S$.
The average component size reads
$\langle s\rangle_S=\sum _{s=1}^Ssp_S(s)\simeq \sum _{s=1}^Ssp(s)=\Lambda(\tau-1,S)/\Lambda(\tau,S)$,
from which $\langle N\rangle_S$ can be
obtained as $\langle N\rangle_S \simeq S/\langle s\rangle _S$.
Using $\Lambda (x,S)\simeq \zeta(x)+S^{1-x}/(1-x)$
for $x\neq0,1$, $\Lambda (1,S)\simeq \ln S$ and $\Lambda (0,S)\simeq S$, valid for large $S$,
we obtain
\begin{equation}
\langle N \rangle_S\simeq
\begin{cases}
      \displaystyle
      (2-\tau)/(1-\tau) &;\, \text{for}\quad \tau<1 \\
      \ln{S} &;\, \text{for}\quad \tau=1 \\
      \zeta(\tau)(2-\tau)S^{\tau-1} &;\, \text{for}\quad 1<\tau<2 \\
      \displaystyle
      \zeta(2)S/\ln{S} &;\, \text{for}\quad \tau=2 \\
      \displaystyle
      \zeta(\tau)S/\zeta(\tau-1) &;\, \text{for}\quad \tau>2 \ ,
    \end{cases}    
    \label{aveNext}
\end{equation}
which is in excellent agreement with the exact determination, see the SM~\cite{Sup}. 
From the definition $\langle N\rangle_S \,\,p_S(s^*)=1$, we obtain
$s^*(\tau,S)\simeq [S/\Lambda(\tau-1,S)]^{1/\tau}$ and, substituting in Eq.~(\ref{JefH}),
one arrives at
the sought after result for the average diversity: 
$\langle D \rangle_S \simeq
s^*+(s^*)^{\tau}[\zeta(\tau,s^*)-\zeta(\tau,S+1)]$.
Approximating the Riemann zeta function by
$\zeta(x)\simeq (x-1)^{-1}+\gamma$, where $\gamma\simeq 0.577$ is the Euler constant, we can write
\begin{eqnarray}
s^*(\tau,S)&\simeq& S^{1/\tau}\left[\gamma+(S^{2-\tau}-1)/(2-\tau)\right]^{-1/\tau} \\
\langle D \rangle_S &\simeq& \frac{\tau s^*-(s^*)^{\tau}S^{1-\tau}}{\tau -1}, \label{eq.finalD}
\end{eqnarray}
where the appropriate limits for $\tau=1$ and 2 are taken.

\begin{figure}
\includegraphics[width=0.9\linewidth]{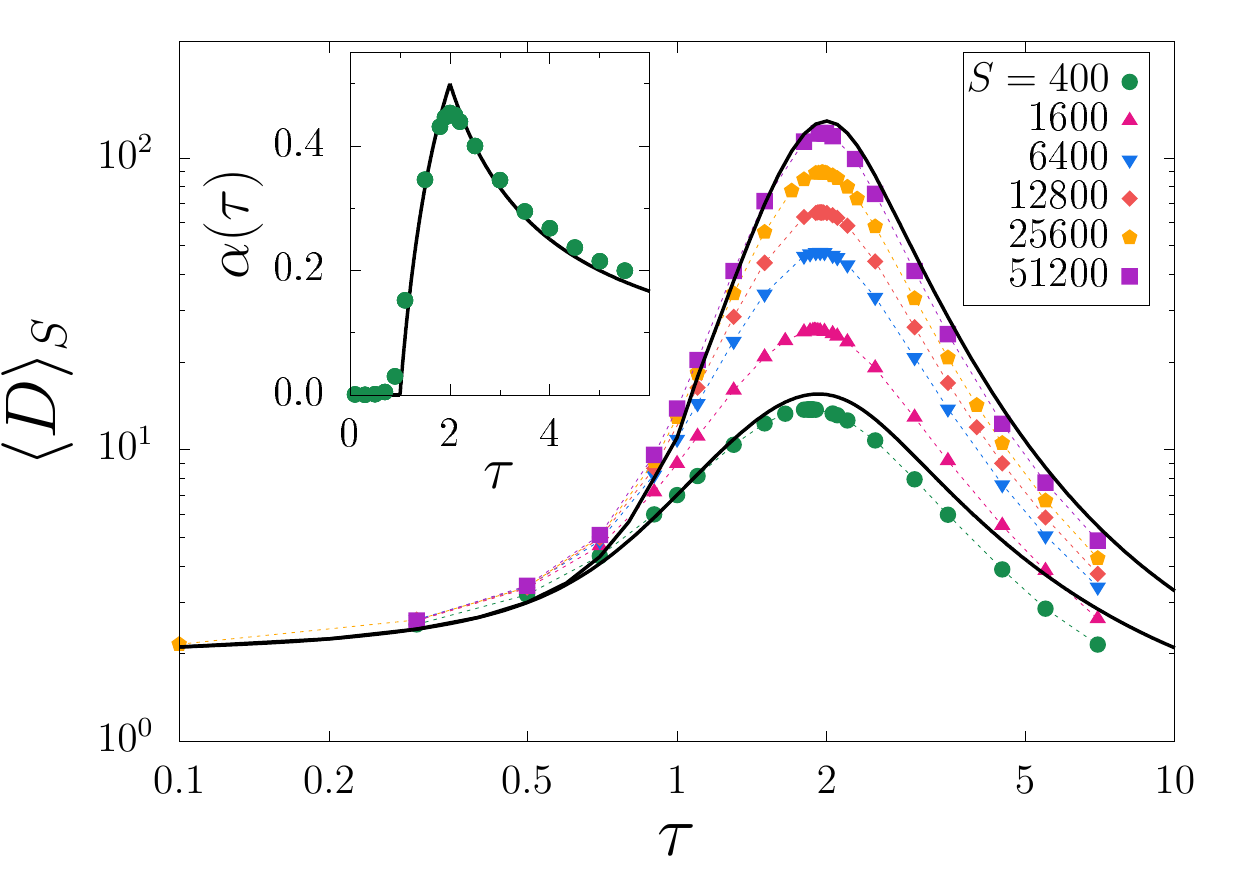}
\caption{Average diversity, $\langle D\rangle _S$, obtained through numerical 
simulations for various sizes $S$ (see key), dashed lines are guides to the eye. 
Entities are extracted from the bare distribution Eq.~(\ref{barealg}) and the statistics
is restricted over configurations respecting the global constraint.
Results are averaged over $10^4$-$10^6$ configurations. Solid lines (shown only for the extreme sizes), are the analytical 
solutions given by Eq.~(\ref{eq.finalD}). Inset: The exponent \(\alpha(\tau)\), defined below Eq.~(\ref{analD}),
as a function of \(\tau\). Solid line is the analytical result, dots are fits from the simulation data.}
\label{pic:D}
\end{figure}

This determination of $\langle D\rangle_S$ is portrayed in Fig.~\ref{pic:D} 
and compared with the outcome of numerical simulations finding a very good agreement.
For large $S$, the leading contribution to Eq.~(\ref{eq.finalD}) is
\begin{equation}
\langle D \rangle_S
\simeq
    \begin{cases}
      \displaystyle
      (2-\tau)/(1-\tau) &; \,\text{for}\quad \tau<1 \\
      \ln S &; \,\text{for}\quad \tau=1 \\
      \displaystyle
      \frac{\tau (2-\tau)^{1/\tau}}{\tau-1}S^{1-1/\tau} &;\, \text{for}\quad 1<\tau<2 \\
      \displaystyle
      2\left(S/\ln S\right)^{1/2} &;\, \text{for}\quad \tau=2 \\
      \displaystyle
      \frac{\tau}{\tau-1}\left[\frac{S}{\gamma+(\tau-2)^{-1}}\right]^{1/\tau} &;\, \text{for}\quad \tau>2 \ .
    \end{cases} 
    \label{analD}
\end{equation}
One has $\langle D\rangle _S\sim S^{\alpha(\tau)}$ with
$\alpha(\tau)=0$ for $\tau<1$, $\alpha(\tau)=1-1/\tau$ for $1<\tau<2$ and $\alpha(\tau)=1/\tau$ for $\tau>2$,
see inset of Fig~\ref{pic:D}.
In conclusion, for large $S$, $\langle D \rangle_S$
presents a pronounced peak at $\tau= 2$.
This behavior is due to the competition between the abundance of entities
$\langle N\rangle_S$, favored by large $\tau$, and the diversity of their sizes which
instead is enhanced by small $\tau$, as shown in Fig.~\ref{pic:pictorial}.
We remark that the upper bound obtained by considering the 
deterministic partition $S\simeq1+2+\ldots+D$ with $D\sim S^{1/2}$
overpowers the $\tau=2$ case only by a logarithmic factor.

Let us mention that, although we explicitly solved the model for power law distributions,
which yield maximum diversity,
our calculations can be straightforwardly generalized to different $p(s)$. 
For instance, in the case of algebraic distributions with a lower cut-off, a case often representative of real situations~\cite{DeMarzo21}, 
one recovers similar results provided that the cut-off is independent of $S$ (see the SM~\cite{Sup}).

We also stress that, as shown in the SM~\cite{Sup}, among the possible measures 
of diversity usually considered in the literature, $D$ is the only one to be maximized in connection with
Zipf's law.

We notice also that the model considered here is related to the random 
allocation model~\cite{Godr_che_2019} where
the resource $S$ is distributed among an {\it assigned} number $N$ of components. The diversity properties of such model,
however, are very different and, in particular, the special role played by $\tau=2$ is missing.
This is briefly discussed in the SM~\cite{Sup}.

{\it Diversity, Zipf's and Heaps' laws.---}
Since the diversity is determined once an empirical distribution of sizes is given,  we can
use $\langle D\rangle_S$ given in Eq.~(\ref{analD})
to estimate the diversity index $D$ of power law distributed empirical data,
regardless of the mechanism whereby they are produced.
If this assumption holds, on the basis of our analytical arguments, one can conclude
that if a system displays Zipf's law 
($\tau\simeq 2$) it is at the edge of maximal diversity and {\it vice versa}.

As a first example we consider quantitative linguistics, the field  in which 
Zipf's law has been originally observed
in almost every human language~\cite{Condon28,Piantadosi14,Gerlach14,MoFoCo16}.
The regime of validity of the 
law in this context~\cite{Font13}, its deviations~\cite{Ferrer05} and 
the underlying mechanism(s) are still a matter of dispute. Nonetheless, large scale
studies have been performed in order to validate that. For example, Moreno-S\'anchez
{\it et al.}~\cite{MoFoCo16} considered a very large set of English books (more than 30000) from the Gutenberg
Project database. They checked how well some simple, one-parameter forms of the
Zipf's law describe these data on the whole interval of frequencies, finding
very good agreement with a distribution of exponents centered on $\tau\simeq 2$.

\begin{figure}
\includegraphics[width=\linewidth]{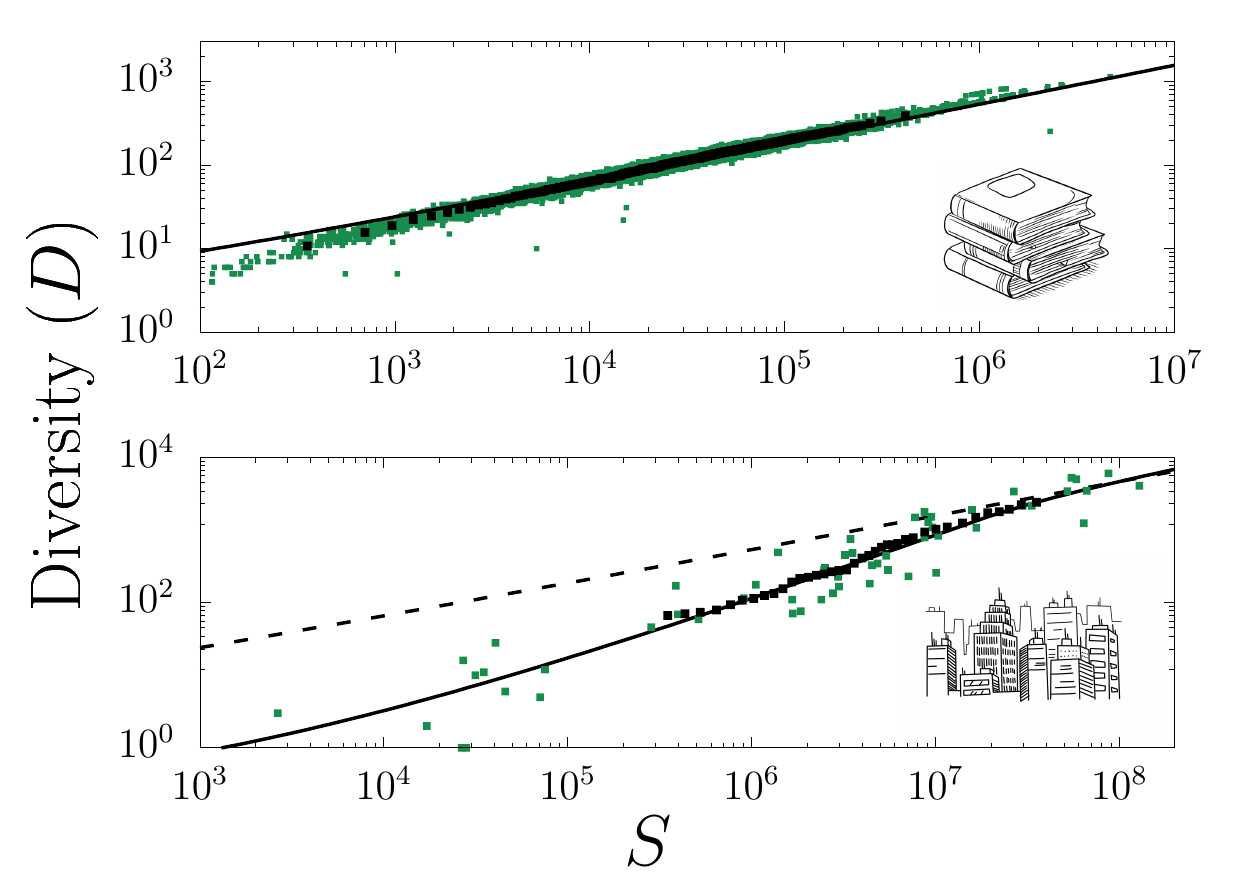}
\caption{(Top panel) Diversity index $D$ evaluated from the data of Ref.~\cite{MoFoCo16}. Each
green point is one of the more than 30000 English books in the Project Gutenberg database (accessed July 2014), 
while the black squares are a running average over 20 points.
The solid line is the result $\langle D\rangle_S=2(S/\ln S)^{1/2}$,
from Eq.~(\ref{analD}) for $\tau=2$, which corresponds to maximal diversity. 
(Bottom panel) Diversity index using data from the GeoNames database~\cite{GeoNames} for
cities. Each green point is an European country and the black squares are the corresponding running average. 
The solid line is Eq.~(29) of the SM~\cite{Sup} where the presence of a lower cutoff $s_L$ is taken
into account. $s_L$ is estimated from the average of the smallest city in each
country ($s_L\simeq1313$), see SM~\cite{Sup}. The dashed line is the behavior $\langle D\rangle_S=2(S/\ln S)^{1/2}$, which is
approached only asymptotically.
}
\label{pic:books}
\end{figure}

We use the filtered data of Ref.~\cite{MoFoCo16} and, for each book,
measure the diversity index $D$. 
The total number of words a book contains is its total size $S$, 
the number of distinct words
is the number of entities, $N$, and the size $s$ of each entity
is its absolute frequency, i.e. how many times that word appears.
The diversity $D$ is therefore the number of {\it different frequencies}
a given text displays.
The result of this analysis is shown in Fig.~\ref{pic:books} along with Eq.~(\ref{analD}) for $\tau=2$.
Notice that there are no free parameters in the plot. 
The agreement between our theoretical prediction and the experimental points is
consistent with the results reported in Ref.~\cite{MoFoCo16} showing that a great deal of the
books have $\tau$ close to 2.

As a second example, we consider how the total population $S$ of a country is 
distributed among its cities. We use data for European countries from the GeoNames database~\cite{GeoNames},
for which Simini and James~\cite{SiJa19} showed that the size $s$ of cities closely follows a Zipf's distribution
($\tau \simeq 2.02$). The diversity index $D$ is shown in Fig.~\ref{pic:books} (bottom panel). 
Since  cities cannot be smaller than a certain lower cutoff $s_L$, the analytical prediction to compare with
is Eq.~(29) of the SM, see SM~\cite{Sup}, (solid line). Despite the noisy character of the data, there is a very good agreement
between the data and our theory.

The content of Eq.~(\ref{aveNext}) is Heaps' law, 
which gives an estimate of the number of components
of a system of total size $S$ given that the empirical size distribution follows
a power law with exponent $\tau$.
Our expression of the law for $\tau>1$ is in accordance with Ref.~\cite{LuZhZh10}
and complements the result with the cases with $\tau \leq 1$ 
and with the appropriate prefactors.  
Heap's law is expected to hold for systems which are robust in the statistics
of their component ($p_S(s)$ in our notation) at varying $S$~\cite{LuZhZh10,DeMarzo21}. 
This is captured in our approach, where Eq.~(\ref{aveNext}) is only arrived at using distributions which have the same form for any $S$ 
(the same applies to Eq.~(\ref{analD})).

In our approach, Heap's law~(\ref{aveNext}) and the diversity law~(\ref{analD}) imply each other,
encoding dependencies on the system size on equal footings.
However, notably, the latter naturally selects the exponent $\tau=2$ as a special one.
Moreover, our analysis of the Gutenberg dataset shows that the diversity law is obeyed up to the largest sizes considered ($S\simeq10^7$),
whereas it is known~\cite{Lu13} that strong deviations from Heaps' law are caused 
by the finiteness of the vocabulary.
Therefore, at least in the context of language, the diversity law appears more robust
and this suggests that its use could be more suited to interpret
the size dependence of empirical data.

{\it Discussion.---} The partition of a finite resource $S$
among constituents informs numerous systems in diverse
fields of science and humanities. 
In this Letter, by solving a paradigmatic statistical model, 
we have shown that a maximally diverse partition is accompanied by Zipf's law.
Such co-occurrence is a general property of the empirical distribution, holding 
irrespectively of the specific mechanisms at work in generating Zipfian behavior
in given systems. 

Diversity and information are fundamental concepts 
for the description of complex statistical systems 
whose formalization led to the definition of 
 a coherent set of quantitative measures, Boltzmann-Shannon entropy
above all. Our results show that in the case of system obeying Zipf's law
an important role is played by one of such measures, the index $D$.
When framed in terms of extremization of appropriate cost functions,
problems are endowed with a complementary description 
and can be approached with new strategies.
Our study suggests that,
in some instances where Zipf's law is empirically observed, 
promoting diversity to the role of a driving force
could provide further theoretical insights towards
a deeper and more general comprehension.
\\\\

O.M. is indebted with I. A. Hatton, M. Smerlak and A. Zadorin for numerous and insightful discussions
and acknowledges the Alexander von Humboldt Foundation 
in the framework of the Sofja Kovalevskaja Award endowed by 
the German Federal Ministry of Education and Research
for providing funding for this work.
A.A.L. and J.J.A. thank Salerno University for hospitality.
A.A.L acknowledges the Brazilian funding agency CAPES in the framework of the Capes-PrInt program (grant 88887.466912/2019-00). 
J.J.A. thanks the Brazilian funding agency CNPq (grant 308927/2017-6).
The authors thank S. Bora for the drawings in Fig.~\ref{pic:books}.

\bibliography{biblioH}

\begin{thebibliography}{47}
\expandafter\ifx\csname natexlab\endcsname\relax\def\natexlab#1{#1}\fi
\expandafter\ifx\csname bibnamefont\endcsname\relax
  \def\bibnamefont#1{#1}\fi
\expandafter\ifx\csname bibfnamefont\endcsname\relax
  \def\bibfnamefont#1{#1}\fi
\expandafter\ifx\csname citenamefont\endcsname\relax
  \def\citenamefont#1{#1}\fi
\expandafter\ifx\csname url\endcsname\relax
  \def\url#1{\texttt{#1}}\fi
\expandafter\ifx\csname urlprefix\endcsname\relax\def\urlprefix{URL }\fi
\providecommand{\bibinfo}[2]{#2}
\providecommand{\eprint}[2][]{\url{#2}}

\bibitem[{\citenamefont{Jost}(2006)}]{Jost06}
\bibinfo{author}{\bibfnamefont{L.}~\bibnamefont{Jost}},
  \bibinfo{journal}{Oikos} \textbf{\bibinfo{volume}{113}}, \bibinfo{pages}{363}
  (\bibinfo{year}{2006}).

\bibitem[{\citenamefont{Tuomisto}(2010)}]{Tuomisto10}
\bibinfo{author}{\bibfnamefont{H.}~\bibnamefont{Tuomisto}},
  \bibinfo{journal}{Oecologia} \textbf{\bibinfo{volume}{164}},
  \bibinfo{pages}{853} (\bibinfo{year}{2010}).

\bibitem[{\citenamefont{Ives and Carpenter}(2007)}]{Ives58}
\bibinfo{author}{\bibfnamefont{A.~R.} \bibnamefont{Ives}} \bibnamefont{and}
  \bibinfo{author}{\bibfnamefont{S.~R.} \bibnamefont{Carpenter}},
  \bibinfo{journal}{Science} \textbf{\bibinfo{volume}{317}},
  \bibinfo{pages}{58} (\bibinfo{year}{2007}).

\bibitem[{\citenamefont{Elton}(1958)}]{Elton1958}
\bibinfo{author}{\bibfnamefont{C.~S.} \bibnamefont{Elton}},
  \emph{\bibinfo{title}{The ecology of invasions by animals and plants}}
  (\bibinfo{publisher}{Methuen \& Co. Ltd.}, \bibinfo{address}{London, UK},
  \bibinfo{year}{1958}).

\bibitem[{\citenamefont{Tilman et~al.}(2006)\citenamefont{Tilman, Reich, and
  Knops}}]{Tilman06}
\bibinfo{author}{\bibfnamefont{D.}~\bibnamefont{Tilman}},
  \bibinfo{author}{\bibfnamefont{P.~B.} \bibnamefont{Reich}}, \bibnamefont{and}
  \bibinfo{author}{\bibfnamefont{J.~M.} \bibnamefont{Knops}},
  \bibinfo{journal}{Nature} \textbf{\bibinfo{volume}{441}},
  \bibinfo{pages}{629} (\bibinfo{year}{2006}).

\bibitem[{\citenamefont{Arese~Lucini et~al.}(2020)\citenamefont{Arese~Lucini,
  Morone, Tomassone, and Makse}}]{AreseLucini20}
\bibinfo{author}{\bibfnamefont{F.}~\bibnamefont{Arese~Lucini}},
  \bibinfo{author}{\bibfnamefont{F.}~\bibnamefont{Morone}},
  \bibinfo{author}{\bibfnamefont{M.~S.} \bibnamefont{Tomassone}},
  \bibnamefont{and} \bibinfo{author}{\bibfnamefont{H.~A.} \bibnamefont{Makse}},
  \bibinfo{journal}{PLOS ONE} \textbf{\bibinfo{volume}{15}}, \bibinfo{pages}{1}
  (\bibinfo{year}{2020}).

\bibitem[{\citenamefont{Tacchella et~al.}(2012)\citenamefont{Tacchella,
  Cristelli, Caldarelli, Gabrielli, and Pietronero}}]{Tacchella12}
\bibinfo{author}{\bibfnamefont{A.}~\bibnamefont{Tacchella}},
  \bibinfo{author}{\bibfnamefont{M.}~\bibnamefont{Cristelli}},
  \bibinfo{author}{\bibfnamefont{G.}~\bibnamefont{Caldarelli}},
  \bibinfo{author}{\bibfnamefont{A.}~\bibnamefont{Gabrielli}},
  \bibnamefont{and}
  \bibinfo{author}{\bibfnamefont{L.}~\bibnamefont{Pietronero}},
  \bibinfo{journal}{Sci. Rep.} \textbf{\bibinfo{volume}{2}},
  \bibinfo{pages}{723} (\bibinfo{year}{2012}).

\bibitem[{\citenamefont{R{\'e}nyi et~al.}(1961)}]{renyi1961measures}
\bibinfo{author}{\bibfnamefont{A.}~\bibnamefont{R{\'e}nyi}}
  \bibnamefont{et~al.}, in \emph{\bibinfo{booktitle}{Proceedings of the Fourth
  Berkeley Symposium on Mathematical Statistics and Probability, Volume 1:
  Contributions to the Theory of Statistics}} (\bibinfo{organization}{The
  Regents of the University of California}, \bibinfo{year}{1961}).

\bibitem[{\citenamefont{Tsallis}(1988)}]{Tsallis88}
\bibinfo{author}{\bibfnamefont{C.}~\bibnamefont{Tsallis}}, \bibinfo{journal}{J.
  Stat. Phys.} \textbf{\bibinfo{volume}{52}}, \bibinfo{pages}{479}
  (\bibinfo{year}{1988}).

\bibitem[{Sup()}]{Sup}
\bibinfo{howpublished}{See Supplemental Material for an account of R\'enyi
  entropies, their connection with diversity indices and arguments for studying
  specifically the diversity index $D$ considered in this paper based on
  numerical simulations, an explicit expression for the probability
  distribution of the diversity $p_S(D)$, an exact computation of the dressed
  probability distribution $p_S(s)$ and $p_S(N)$, motivations for the
  approximation $p_S(s)\simeq p(s)$, the case of power law bare distributions
  with a lower cut-off $s_L$ and details of the analysis of population datasets
  and an account of the behaviour of diversity in the random allocation model.
  Ref.~\cite{PhysRevE.95.032136} is included.}

\bibitem[{\citenamefont{Gabaix}(1999)}]{Gabaix99}
\bibinfo{author}{\bibfnamefont{X.}~\bibnamefont{Gabaix}}, \bibinfo{journal}{The
  Quarterly Journal of Economics} \textbf{\bibinfo{volume}{114}},
  \bibinfo{pages}{739} (\bibinfo{year}{1999}).

\bibitem[{\citenamefont{Piantadosi}(2014)}]{Piantadosi14}
\bibinfo{author}{\bibfnamefont{S.~T.} \bibnamefont{Piantadosi}},
  \bibinfo{journal}{Psychon Bull Rev.} \textbf{\bibinfo{volume}{21}},
  \bibinfo{pages}{1112} (\bibinfo{year}{2014}).

\bibitem[{\citenamefont{Furusawa and Kaneko}(2003)}]{Furusawa03}
\bibinfo{author}{\bibfnamefont{C.}~\bibnamefont{Furusawa}} \bibnamefont{and}
  \bibinfo{author}{\bibfnamefont{K.}~\bibnamefont{Kaneko}},
  \bibinfo{journal}{Phys. Rev. Lett.} \textbf{\bibinfo{volume}{90}},
  \bibinfo{pages}{088102} (\bibinfo{year}{2003}).

\bibitem[{\citenamefont{Zipf}(1949)}]{Zipf49}
\bibinfo{author}{\bibfnamefont{G.~K.} \bibnamefont{Zipf}},
  \emph{\bibinfo{title}{Human Behaviour and the Principle of Least Effort: An
  Introduction to Human Ecology}} (\bibinfo{publisher}{Addison-Wesley,
  Cambridge, MA}, \bibinfo{year}{1949}).

\bibitem[{\citenamefont{Newman}(2005)}]{Newman05}
\bibinfo{author}{\bibfnamefont{M.~E.~J.} \bibnamefont{Newman}},
  \bibinfo{journal}{Contemp. Phys.} \textbf{\bibinfo{volume}{46}},
  \bibinfo{pages}{323} (\bibinfo{year}{2005}).

\bibitem[{\citenamefont{Clauset et~al.}(2009)\citenamefont{Clauset, Shalizi,
  and Newman}}]{Clauset09}
\bibinfo{author}{\bibfnamefont{A.}~\bibnamefont{Clauset}},
  \bibinfo{author}{\bibfnamefont{C.~R.} \bibnamefont{Shalizi}},
  \bibnamefont{and} \bibinfo{author}{\bibfnamefont{M.~E.}
  \bibnamefont{Newman}}, \bibinfo{journal}{SIAM Review}
  \textbf{\bibinfo{volume}{51}}, \bibinfo{pages}{661} (\bibinfo{year}{2009}).

\bibitem[{\citenamefont{Cristelli et~al.}(2012)\citenamefont{Cristelli, Batty,
  and Pietronero}}]{Cristelli12}
\bibinfo{author}{\bibfnamefont{M.}~\bibnamefont{Cristelli}},
  \bibinfo{author}{\bibfnamefont{M.}~\bibnamefont{Batty}}, \bibnamefont{and}
  \bibinfo{author}{\bibfnamefont{L.}~\bibnamefont{Pietronero}},
  \bibinfo{journal}{Sci. Rep.} \textbf{\bibinfo{volume}{2}},
  \bibinfo{pages}{812} (\bibinfo{year}{2012}).

\bibitem[{\citenamefont{Axtell}(2001)}]{Axtell01}
\bibinfo{author}{\bibfnamefont{R.~L.} \bibnamefont{Axtell}},
  \bibinfo{journal}{Science} \textbf{\bibinfo{volume}{293}},
  \bibinfo{pages}{1818} (\bibinfo{year}{2001}).

\bibitem[{\citenamefont{Willis and Yule}(1922)}]{Willis22}
\bibinfo{author}{\bibfnamefont{J.~C.} \bibnamefont{Willis}} \bibnamefont{and}
  \bibinfo{author}{\bibfnamefont{G.~U.} \bibnamefont{Yule}},
  \bibinfo{journal}{Nature} \textbf{\bibinfo{volume}{109}},
  \bibinfo{pages}{177} (\bibinfo{year}{1922}).

\bibitem[{\citenamefont{Oddershede et~al.}(1993)\citenamefont{Oddershede,
  Dimon, and Bohr}}]{OdDiBo93}
\bibinfo{author}{\bibfnamefont{L.}~\bibnamefont{Oddershede}},
  \bibinfo{author}{\bibfnamefont{P.}~\bibnamefont{Dimon}}, \bibnamefont{and}
  \bibinfo{author}{\bibfnamefont{J.}~\bibnamefont{Bohr}},
  \bibinfo{journal}{Phys. Rev. Lett.} \textbf{\bibinfo{volume}{71}},
  \bibinfo{pages}{3107} (\bibinfo{year}{1993}).

\bibitem[{\citenamefont{Corral et~al.}(2020)\citenamefont{Corral, Serra, and
  {Ferrer-i-Cancho}}}]{Corral}
\bibinfo{author}{\bibfnamefont{A.}~\bibnamefont{Corral}},
  \bibinfo{author}{\bibfnamefont{I.}~\bibnamefont{Serra}}, \bibnamefont{and}
  \bibinfo{author}{\bibfnamefont{R.}~\bibnamefont{{Ferrer-i-Cancho}}},
  \bibinfo{journal}{Phys. Rev. E} \textbf{\bibinfo{volume}{102}},
  \bibinfo{pages}{052113} (\bibinfo{year}{2020}).

\bibitem[{\citenamefont{Simon}(1955)}]{Simon55}
\bibinfo{author}{\bibfnamefont{H.~A.} \bibnamefont{Simon}},
  \bibinfo{journal}{Biometrika} \textbf{\bibinfo{volume}{42}},
  \bibinfo{pages}{425} (\bibinfo{year}{1955}).

\bibitem[{\citenamefont{Levy and Solomon}(1996)}]{Levy96}
\bibinfo{author}{\bibfnamefont{M.}~\bibnamefont{Levy}} \bibnamefont{and}
  \bibinfo{author}{\bibfnamefont{S.}~\bibnamefont{Solomon}},
  \bibinfo{journal}{Int. J. Mod. Phys. {C}} \textbf{\bibinfo{volume}{7}},
  \bibinfo{pages}{595} (\bibinfo{year}{1996}).

\bibitem[{\citenamefont{Marsili and Zhang}(1998)}]{Marsili98}
\bibinfo{author}{\bibfnamefont{M.}~\bibnamefont{Marsili}} \bibnamefont{and}
  \bibinfo{author}{\bibfnamefont{Y.-C.} \bibnamefont{Zhang}},
  \bibinfo{journal}{Phys. Rev. Lett.} \textbf{\bibinfo{volume}{80}},
  \bibinfo{pages}{2741} (\bibinfo{year}{1998}).

\bibitem[{\citenamefont{{Ferrer-i-Cancho} and Sol{\'e}}(2003)}]{Cancho03}
\bibinfo{author}{\bibfnamefont{R.}~\bibnamefont{{Ferrer-i-Cancho}}}
  \bibnamefont{and} \bibinfo{author}{\bibfnamefont{R.~V.}
  \bibnamefont{Sol{\'e}}}, \bibinfo{journal}{Proceedings of the National
  Academy of Sciences} \textbf{\bibinfo{volume}{100}}, \bibinfo{pages}{788}
  (\bibinfo{year}{2003}).

\bibitem[{\citenamefont{Tria et~al.}(2014)\citenamefont{Tria, Loreto, Servedio,
  and Strogatz}}]{TrLoSeSt14}
\bibinfo{author}{\bibfnamefont{F.}~\bibnamefont{Tria}},
  \bibinfo{author}{\bibfnamefont{V.}~\bibnamefont{Loreto}},
  \bibinfo{author}{\bibfnamefont{V.~D.~P.} \bibnamefont{Servedio}},
  \bibnamefont{and} \bibinfo{author}{\bibfnamefont{S.~H.}
  \bibnamefont{Strogatz}}, \bibinfo{journal}{Sci. Rep.}
  \textbf{\bibinfo{volume}{4}}, \bibinfo{pages}{5890} (\bibinfo{year}{2014}).

\bibitem[{\citenamefont{Corominas-Murtra
  et~al.}(2015)\citenamefont{Corominas-Murtra, Hanel, and
  Thurner}}]{Corominas15}
\bibinfo{author}{\bibfnamefont{B.}~\bibnamefont{Corominas-Murtra}},
  \bibinfo{author}{\bibfnamefont{R.}~\bibnamefont{Hanel}}, \bibnamefont{and}
  \bibinfo{author}{\bibfnamefont{S.}~\bibnamefont{Thurner}},
  \bibinfo{journal}{Proceedings of the National Academy of Sciences}
  \textbf{\bibinfo{volume}{112}}, \bibinfo{pages}{5348} (\bibinfo{year}{2015}).

\bibitem[{\citenamefont{Mazzolini et~al.}(2018)\citenamefont{Mazzolini,
  Gherardi, Caselle, Lagomarsino, and Osella}}]{Mazzolini18}
\bibinfo{author}{\bibfnamefont{A.}~\bibnamefont{Mazzolini}},
  \bibinfo{author}{\bibfnamefont{M.}~\bibnamefont{Gherardi}},
  \bibinfo{author}{\bibfnamefont{M.}~\bibnamefont{Caselle}},
  \bibinfo{author}{\bibfnamefont{M.~C.} \bibnamefont{Lagomarsino}},
  \bibnamefont{and} \bibinfo{author}{\bibfnamefont{M.}~\bibnamefont{Osella}},
  \bibinfo{journal}{Phys. Rev. X} \textbf{\bibinfo{volume}{8}},
  \bibinfo{pages}{021023} (\bibinfo{year}{2018}).

\bibitem[{\citenamefont{Mora and Bialek}(2011)}]{Mora11}
\bibinfo{author}{\bibfnamefont{T.}~\bibnamefont{Mora}} \bibnamefont{and}
  \bibinfo{author}{\bibfnamefont{W.}~\bibnamefont{Bialek}},
  \bibinfo{journal}{J. Stat. Phys.} \textbf{\bibinfo{volume}{144}},
  \bibinfo{pages}{268} (\bibinfo{year}{2011}).

\bibitem[{\citenamefont{Marsili et~al.}(2013)\citenamefont{Marsili,
  Mastromatteo, and Roudi}}]{Marsili13}
\bibinfo{author}{\bibfnamefont{M.}~\bibnamefont{Marsili}},
  \bibinfo{author}{\bibfnamefont{I.}~\bibnamefont{Mastromatteo}},
  \bibnamefont{and} \bibinfo{author}{\bibfnamefont{Y.}~\bibnamefont{Roudi}},
  \bibinfo{journal}{J. Stat. Mech.: Theory and Experiment}
  \textbf{\bibinfo{volume}{2013}}, \bibinfo{pages}{P09003}
  (\bibinfo{year}{2013}).

\bibitem[{\citenamefont{Schwab et~al.}(2014)\citenamefont{Schwab, Nemenman, and
  Mehta}}]{Schwab14}
\bibinfo{author}{\bibfnamefont{D.~J.} \bibnamefont{Schwab}},
  \bibinfo{author}{\bibfnamefont{I.}~\bibnamefont{Nemenman}}, \bibnamefont{and}
  \bibinfo{author}{\bibfnamefont{P.}~\bibnamefont{Mehta}},
  \bibinfo{journal}{Phys. Rev. Lett.} \textbf{\bibinfo{volume}{113}},
  \bibinfo{pages}{068102} (\bibinfo{year}{2014}).

\bibitem[{\citenamefont{Cubero et~al.}(2019)\citenamefont{Cubero, Jo, Marsili,
  Roudi, and Song}}]{Cubero_2019}
\bibinfo{author}{\bibfnamefont{R.~J.} \bibnamefont{Cubero}},
  \bibinfo{author}{\bibfnamefont{J.}~\bibnamefont{Jo}},
  \bibinfo{author}{\bibfnamefont{M.}~\bibnamefont{Marsili}},
  \bibinfo{author}{\bibfnamefont{Y.}~\bibnamefont{Roudi}}, \bibnamefont{and}
  \bibinfo{author}{\bibfnamefont{J.}~\bibnamefont{Song}}, \bibinfo{journal}{J.
  Stat. Mech.: Theory and Experiment} \textbf{\bibinfo{volume}{2019}},
  \bibinfo{pages}{063402} (\bibinfo{year}{2019}).

\bibitem[{Pro()}]{ProjectGutemberg}
\bibinfo{howpublished}{Project Gutenberg, \url{www.gutenberg.org}}.

\bibitem[{Geo()}]{GeoNames}
\bibinfo{howpublished}{GeoNames, \url{www.geonames.org}}.

\bibitem[{\citenamefont{Heaps}(1978)}]{Heaps78}
\bibinfo{author}{\bibfnamefont{H.~S.} \bibnamefont{Heaps}},
  \emph{\bibinfo{title}{Information Retrieval: Computational and Theoretical
  Aspects}} (\bibinfo{publisher}{Academic Press, Inc.},
  \bibinfo{address}{Orlando, FL}, \bibinfo{year}{1978}).

\bibitem[{\citenamefont{L{\"u} et~al.}(2010)\citenamefont{L{\"u}, Zhang, and
  Zhou}}]{LuZhZh10}
\bibinfo{author}{\bibfnamefont{L.}~\bibnamefont{L{\"u}}},
  \bibinfo{author}{\bibfnamefont{Z.-K.} \bibnamefont{Zhang}}, \bibnamefont{and}
  \bibinfo{author}{\bibfnamefont{T.}~\bibnamefont{Zhou}},
  \bibinfo{journal}{PLOS ONE} \textbf{\bibinfo{volume}{5}},
  \bibinfo{pages}{e14139} (\bibinfo{year}{2010}).

\bibitem[{\citenamefont{de~Azevedo-Lopes
  et~al.}(2020)\citenamefont{de~Azevedo-Lopes, de~la Rocha, de~Oliveira, and
  Arenzon}}]{AzRoOlAr20}
\bibinfo{author}{\bibfnamefont{A.}~\bibnamefont{de~Azevedo-Lopes}},
  \bibinfo{author}{\bibfnamefont{A.~R.} \bibnamefont{de~la Rocha}},
  \bibinfo{author}{\bibfnamefont{P.~M.~C.} \bibnamefont{de~Oliveira}},
  \bibnamefont{and} \bibinfo{author}{\bibfnamefont{J.~J.}
  \bibnamefont{Arenzon}}, \bibinfo{journal}{Phys. Rev. E}
  \textbf{\bibinfo{volume}{101}}, \bibinfo{pages}{012108}
  (\bibinfo{year}{2020}).

\bibitem[{\citenamefont{De~Marzo et~al.}(2021)\citenamefont{De~Marzo,
  Gabrielli, Zaccaria, and Pietronero}}]{DeMarzo21}
\bibinfo{author}{\bibfnamefont{G.}~\bibnamefont{De~Marzo}},
  \bibinfo{author}{\bibfnamefont{A.}~\bibnamefont{Gabrielli}},
  \bibinfo{author}{\bibfnamefont{A.}~\bibnamefont{Zaccaria}}, \bibnamefont{and}
  \bibinfo{author}{\bibfnamefont{L.}~\bibnamefont{Pietronero}},
  \bibinfo{journal}{Phys. Rev. Research} \textbf{\bibinfo{volume}{3}},
  \bibinfo{pages}{013084} (\bibinfo{year}{2021}).

\bibitem[{\citenamefont{Godr{\`{e}}che}(2019)}]{Godr_che_2019}
\bibinfo{author}{\bibfnamefont{C.}~\bibnamefont{Godr{\`{e}}che}},
  \bibinfo{journal}{Journal of Statistical Mechanics: Theory and Experiment}
  \textbf{\bibinfo{volume}{2019}}, \bibinfo{pages}{063207}
  (\bibinfo{year}{2019}).

\bibitem[{\citenamefont{Condon}(1928)}]{Condon28}
\bibinfo{author}{\bibfnamefont{E.~U.} \bibnamefont{Condon}},
  \bibinfo{journal}{Science} \textbf{\bibinfo{volume}{67}},
  \bibinfo{pages}{300} (\bibinfo{year}{1928}).

\bibitem[{\citenamefont{Gerlach and Altmann}(2014)}]{Gerlach14}
\bibinfo{author}{\bibfnamefont{M.}~\bibnamefont{Gerlach}} \bibnamefont{and}
  \bibinfo{author}{\bibfnamefont{E.~G.} \bibnamefont{Altmann}},
  \bibinfo{journal}{New J. Phys.} \textbf{\bibinfo{volume}{16}},
  \bibinfo{pages}{113010} (\bibinfo{year}{2014}).

\bibitem[{\citenamefont{Moreno-S\'anchez
  et~al.}(2016)\citenamefont{Moreno-S\'anchez, Font-Clos, and
  Corral}}]{MoFoCo16}
\bibinfo{author}{\bibfnamefont{I.}~\bibnamefont{Moreno-S\'anchez}},
  \bibinfo{author}{\bibfnamefont{F.}~\bibnamefont{Font-Clos}},
  \bibnamefont{and} \bibinfo{author}{\bibfnamefont{A.}~\bibnamefont{Corral}},
  \bibinfo{journal}{PLOS ONE} \textbf{\bibinfo{volume}{11}}, \bibinfo{pages}{1}
  (\bibinfo{year}{2016}).

\bibitem[{\citenamefont{Font-Clos et~al.}(2013)\citenamefont{Font-Clos, Boleda,
  and Corral}}]{Font13}
\bibinfo{author}{\bibfnamefont{F.}~\bibnamefont{Font-Clos}},
  \bibinfo{author}{\bibfnamefont{G.}~\bibnamefont{Boleda}}, \bibnamefont{and}
  \bibinfo{author}{\bibfnamefont{A.}~\bibnamefont{Corral}},
  \bibinfo{journal}{New J. Phys.} \textbf{\bibinfo{volume}{15}},
  \bibinfo{pages}{093033} (\bibinfo{year}{2013}).

\bibitem[{\citenamefont{{Ferrer-i-Cancho}}(2005)}]{Ferrer05}
\bibinfo{author}{\bibfnamefont{R.}~\bibnamefont{{Ferrer-i-Cancho}}},
  \bibinfo{journal}{Eur. Phys. J. B} \textbf{\bibinfo{volume}{44}},
  \bibinfo{pages}{249} (\bibinfo{year}{2005}).

\bibitem[{\citenamefont{Simini and James}(2019)}]{SiJa19}
\bibinfo{author}{\bibfnamefont{F.}~\bibnamefont{Simini}} \bibnamefont{and}
  \bibinfo{author}{\bibfnamefont{C.}~\bibnamefont{James}},
  \bibinfo{journal}{EPJ Data Science} \textbf{\bibinfo{volume}{8}},
  \bibinfo{pages}{24} (\bibinfo{year}{2019}).

\bibitem[{\citenamefont{L{\"u} et~al.}(2013)\citenamefont{L{\"u}, Zhang, and
  Zhou}}]{Lu13}
\bibinfo{author}{\bibfnamefont{L.}~\bibnamefont{L{\"u}}},
  \bibinfo{author}{\bibfnamefont{Z.-K.} \bibnamefont{Zhang}}, \bibnamefont{and}
  \bibinfo{author}{\bibfnamefont{T.}~\bibnamefont{Zhou}},
  \bibinfo{journal}{Sci. Rep.} \textbf{\bibinfo{volume}{3}},
  \bibinfo{pages}{1082} (\bibinfo{year}{2013}).

\bibitem[{\citenamefont{Corberi}(2017)}]{PhysRevE.95.032136}
\bibinfo{author}{\bibfnamefont{F.}~\bibnamefont{Corberi}},
  \bibinfo{journal}{Phys. Rev. E} \textbf{\bibinfo{volume}{95}},
  \bibinfo{pages}{032136} (\bibinfo{year}{2017}).

\end{thebibliography}

\end{document}